\documentclass[%
superscriptaddress,
 amsmath,amssymb,
 aps,
 pra,
 10pt,
 twocolumn,
 floats,
floatfix,
]{revtex4}

\usepackage[latin1]{inputenc}
\usepackage[T1]{fontenc}
\usepackage{graphicx}
\usepackage{dcolumn}
\usepackage{bm}
\usepackage{units}
\usepackage[squaren]{SIunits}
\usepackage{color}

\newcommand{\ie}{\emph{i.e.}}
\newcommand{\eg}{\emph{e.g.}}

\newcommand{\tRam}{t_\mathrm{Ram}}
\newcommand{\tEcho}{t_\mathrm{E}}

\newcommand{\ttwo}{\mathrm{T}_2^\prime}
\newcommand{\ttwos}{\mathrm{T}_2^\ast}

\newcommand{\ket}[1]{ \left|#1\right\rangle}

\newcommand{\MW}{\mrm{MW}}

\newcommand{\Gauss}{\mathrm{G}}
\newcommand{\dB}{\mrm{dB}}

\newcommand{\mrm}[1]{\mathrm{#1}}

\newcommand{\kcirc}[1]{\ket{#1 \mathrm{C}}}
\newcommand{\kell}[1]{\ket{#1 \mathrm{E}}}

\newcommand{\vv}{\boldsymbol}

\begin{document}

\title{Long-lived circular Rydberg states of laser-cooled Rubidium atoms in a cryostat}

\author{T. Cantat-Moltrecht}
\altaffiliation{These authors equally contributed to this work.}
\author{R. Corti\~nas}
\altaffiliation{These authors equally contributed to this work.}
\author{B. Ravon}
\author{P. M\'ehaignerie}
\author{S. Haroche}
\author{J. M. Raimond}
\author{M. Favier}
\author{M. Brune}
\author{C. Sayrin}

\altaffiliation{Corresponding author}
\email{clement.sayrin@lkb.ens.fr}
\affiliation{%
 Laboratoire Kastler Brossel, Coll\`ege de France, CNRS, ENS-PSL University, Sorbonne-Universit\'e  \\
 11 place Marcelin Berthelot, 75005 Paris, France
}%

\date{\today}

\begin{abstract}

The exquisite properties of Rydberg levels make them particularly appealing for emerging quantum technologies. The lifetime of low-angular-momentum laser-accessible levels is however limited to a few $100\,\micro\second$ by optical transitions and microwave blackbody radiation (BBR) induced transfers at room temperature. A considerable improvement would be obtained with the few $10\,\milli\second$ lifetime of circular Rydberg levels in a cryogenic environment reducing the BBR temperature. We demonstrate the preparation of long-lived circular Rydberg levels of laser-cooled Rubidium atoms in a cryostat. We observe a $3.7\,\milli\second$ lifetime for the circular level of principal quantum number $n=52$. By monitoring the transfers between adjacent circular levels, we estimate \emph{in situ} the microwave BBR temperature to be $(11\pm 2)\,\kelvin$. The measured atomic coherence time ($270\,\micro\second$) is limited here only by technical magnetic field fluctuations. This work opens interesting perspectives for quantum simulation and sensing with cold circular Rydberg atoms.

\end{abstract}

\maketitle


Rydberg atoms, \ie\ atoms prepared in levels with a high principal quantum number $n$, are the focus of a renewed interest~\cite{Saffman2010, Browaeys2020}. Their large coupling to electromagnetic fields, their large mutual dipole-dipole interactions and their long lifetimes make them ideally suited to cavity quantum electrodynamics~\cite{Haroche2013, Seiler2011}, quantum sensing~\cite{Sedlacek2012a, Facon2016, Cox2018}, quantum information~\cite{Saffman2010, Graham2019c} and quantum simulations of spin systems~\cite{Weimer2010, Labuhn2016, Bernien2017, Zeiher2017, Browaeys2020}. However, the laser-accessible low-orbital-angular-momentum ($\ell$) Rydberg levels lifetime, mainly determined by optical transitions, limits evolution and measurement times to a few $100\,\micro\second$. 

The circular Rydberg levels $\kcirc{n}$, with maximum orbital angular momentum ($\ell = |m| = n-1$), have a much longer natural lifetime, of $30\,\milli\second$ for $n\approx 50$. It can even be increased by several orders of magnitude inside a spontaneous-emission inhibiting structure~\cite{Hulet1985}. Moreover, circular Rydberg atoms (CRAs) are immune to photoionization, in contrast to low-$\ell$ Rydberg levels~\cite{Saffman2005}. Thus, they can be efficiently laser-trapped over long times~\cite{Cortinas2019}. All these features make it possible to reach long interaction times with microwave cavities~\cite{Assemat2019}, enabling powerful manipulations of the field quantum state in cavity QED~\cite{Raimond2010}. Longer interrogation times in quantum sensing can provide higher sensitivities to the probed fields~\cite{Dietsche2019}. Longer simulation times would allow quantum simulators to explore slow dynamics, \eg, thermalization~\cite{Nguyen2018}.

All experiments with Rydberg atoms, however, must face the problem of population transfer induced by resonant microwave blackbody radiation (BBR). At room temperature, the high number of BBR photons per mode (100 photons at $50\,\giga\hertz$) significantly reduces the Rydberg levels lifetimes, particularly for the circular ones (by two orders of magnitude for $\kcirc{52}$). Moreover, for low-$\ell$ levels, spurious transfers to nearby states create impurities in the atomic system, detrimental to experiments dressing ground-state atoms with Rydberg levels~\cite{Goldschmidt2016, Zeiher2016}. Fully harnessing the long lifetimes of Rydberg levels thus requires a cryogenic environment.

Here, we demonstrate the preparation in a $4\,\kelvin$ cryostat of the $n=52$ circular Rydberg state from laser-cooled Rubidium atoms at a $10\,\micro\kelvin$ temperature. We measure a $(3.7\pm 0.1)\,\milli\second$ lifetime for $\kcirc{52}$. By monitoring the population transfers between adjacent circular levels, we estimate \emph{in situ} the microwave BBR temperature to be $(11\pm2)\,\kelvin$. We also measure the coherence time of a microwave transition between circular levels and identify its current purely technical limitations. 

\begin{figure}[!t]
\centering
	\includegraphics[width = 0.95\columnwidth]{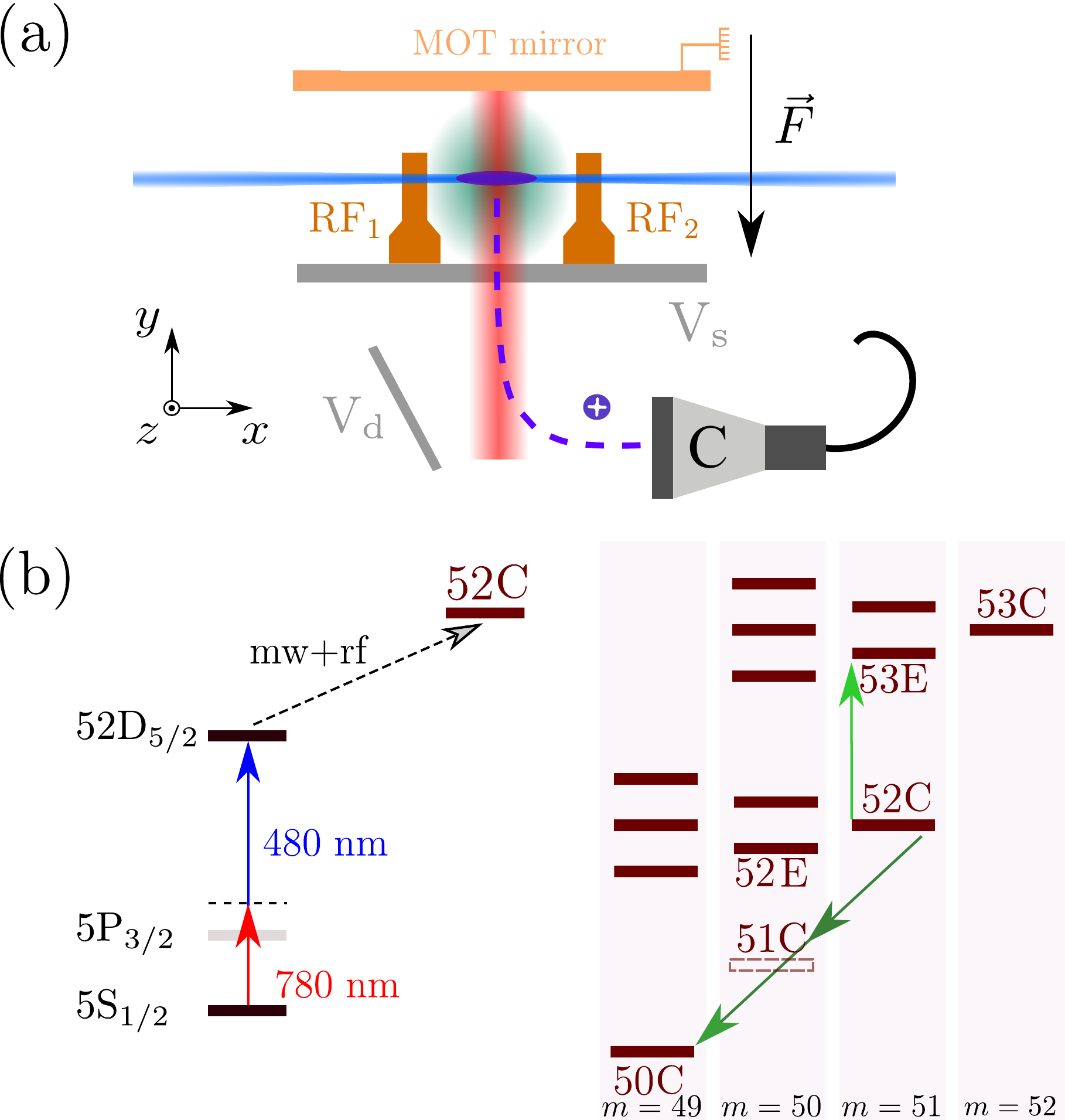}
	\caption{{\bf a.} Experimental setup with axes definition. The blue and red excitation laser beams cross in the cold atom cloud (green) $3\,\milli\meter$ from the surface of the MOT mirror. The electrode $\mrm{V}_\mrm{S}$ applies the electric field $F$ and the ionization field. The Rb$^+$ ions (violet dashed line) are guided to the channeltron C by the deflection electrodes $\mrm{V}_\mrm{d}$. Four additional compensation electrodes (only two, RF1 and RF2, are shown) control electric field gradients and apply the circular state preparation RF field. {\bf b.} Simplified level scheme. Left: two-photon laser excitation (solid arrows) and circular state preparation (dotted arrow). Right: Partial diagram of the Stark levels sorted by $m$ values. The green arrows indicate the one- and two-photon transitions used in the coherence measurements.} 
\label{fig:setup} 
\end{figure}

The experimental setup~\cite{Teixeira2015, Cortinas2019}, sketched on Fig.~\ref{fig:setup}, is enclosed in a wet $4\,\kelvin$ $^4$He optical cryostat. Room-temperature microwave radiation enters the cryostat through SF-57 optical ports (total surface $16\,\centi\meter^2$). In order to keep the microwave BBR temperature as low as possible, we have installed $355\,\centi\meter^2$ of RAM (radar absorption material) plates~\cite{RAM} inside the $4\,\kelvin$ copper thermal shield. From a simple balance of the respective surfaces of RAM and optical ports, we get a rough a priori estimate of the BBR temperature of $17\,\kelvin$.

 Rubidium-87 atoms are laser-cooled and trapped in a 3D mirror magneto-optical trap (MOT) created in front of a Rubidium-coated metallic mirror~\cite{Hermann-Avigliano2014}. An atomic beam that streams out of a 2D MOT along the $z$-axis (axes definition in Fig.~\ref{fig:setup}) loads the 3D MOT. Atoms are further cooled down to about $10\,\micro\kelvin$ via an optical molasses stage that optically pumps them into the $\ket{5\mrm{S}_{1/2}, F = 2}$ ground state.

We excite the atoms from $\ket{5\mrm{S}_{1/2}, F = 2, m_F = 2}$ into $\ket{52\mrm{D}_{5/2}, m_J = 5/2}$ at time $t=0$ by a two-photon laser-excitation (780 and $480\,\nano\meter$ wavelengths). The $780\,\nano\meter$-wavelength red laser beam, blue-detuned from $\ket{5\mrm{P}_{3/2}, F'=3, m_F = 3}$ by $560\,\mega\hertz$, has a waist of $200\,\micro\meter$ and is perpendicular to the mirror surface. It crosses the $480\,\nano\meter$-wavelength blue laser beam ($20\,\micro\meter$ waist), that propagates along the $x$-axis, about $3\,\milli\meter$ away from the mirror surface. Both lasers are pulsed ($5\,\micro\second$ pulse duration).

The excitation to the low-$\ell$ Rydberg level $\ket{52\mrm{D}_{5/2}, m_J = 5/2}$ takes place in a $F = 0.36\,\volt\per\centi\meter$ electric field lifting the degeneracy between the $m_J$ levels. The electric field, $\vv{F}$, is aligned along the $y$-axis and is defined by the electrode $\mrm{V}_\mrm{S}$ and the grounded MOT metallic mirror. Four compensation electrodes, arranged on a $2\,\centi\meter$-side square surrounding the Rydberg-excitation region, are used to minimize the electric field gradients, measured by microwave spectroscopy.

We then efficiently transfer the atoms into the $\ket{52\mrm{F}, m = 2}$ level in an adiabatic process, starting at $t=14\,\micro\second$. We shine a $4\,\micro\second$-long $56.85\,\giga\hertz$-frequency microwave pulse while raising the electric field up to $F = 1.8\,\volt\per\centi\meter$. The microwave is on resonance with the $\ket{52\mrm{D}_{5/2}, m_J = 5/2}\to\ket{52\mrm{F}, m = 2}$ transition when $F=1.75\,\volt\per\centi\meter$. 
The preparation of $\kcirc{52}$ ($n=52, m = +51$) then requires the transfer of 49 orbital momentum quanta. At $t=25\micro\second$, we raise the electric field to $F=2.4\,\volt\per\centi\meter$ and turn on a $\sigma^+$-polarized radio-frequency (RF) field at $\nu_\mrm{RF} = 230\,\mega\hertz$. It is on resonance with the $m\to m+1$ transitions in the Stark manifold when $F = 2.3\,\volt\per\centi\meter$. By scanning the electric field down to $F=2.1\,\volt\per\centi\meter$ in $2\,\micro\second$, we perform a second adiabatic transfer from $\ket{52\mrm{F}, m = 2}$ to $\kcirc{52}$~\cite{Signoles2017}. 
The RF field is produced by applying electric potentials at $\nu_\mrm{RF}$ on two adjacent compensation electrodes. The precise tuning of the amplitudes and relative phase of these potentials makes it possible to cancel the $\sigma^-$ RF polarization component at the position of the atoms. 
Once $\kcirc{52}$ is prepared, we reduce electric field to $F=1.5\,\volt\per\centi\meter$ at $t=34\,\micro\second$.

We finally measure the population of individual Rydberg levels by state-selective field ionization. We apply with electrode $\mrm{V}_\mrm{S}$ a $150\,\micro\second$-long electric field ramp that successively ionizes the Rydberg levels. The $\mathrm{Rb}^+$ ions are then guided to a channeltron using the deflection electrode $\mathrm{V}_\mathrm{d}$ (Fig.\ref{fig:setup})~\cite{Teixeira2015}. By recording the arrival times of the ions, we recover the ionization spectrum of the Rydberg atom cloud. The complete sequence is repeated several hundred times on the same initial cold atom cloud, for a total duration of $\approx1\,\second$ during which the optical molasses is kept on. 

\begin{figure}[t]
\centering
	\includegraphics[width = 0.95\columnwidth]{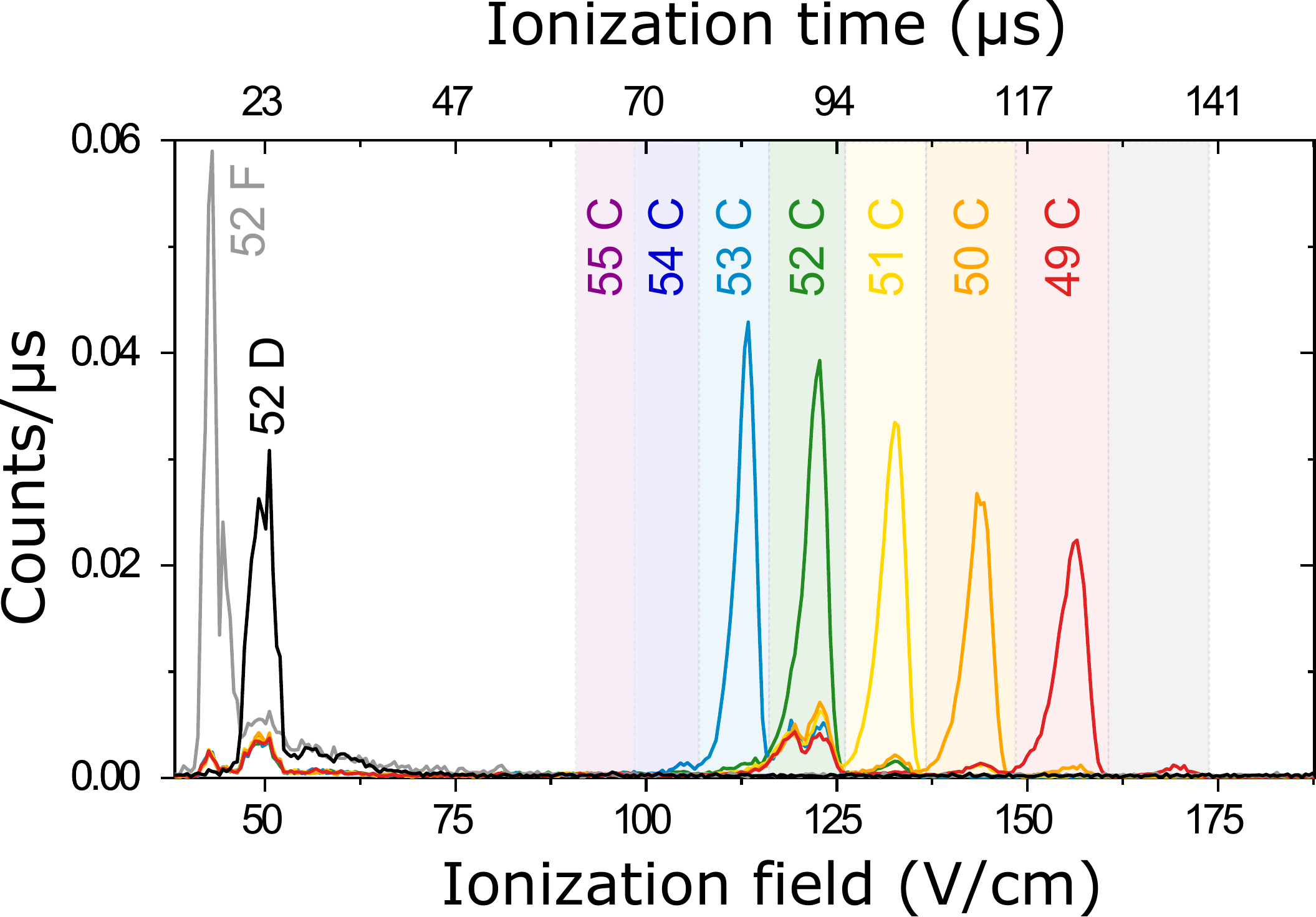}
	\caption{Ionization spectra of two low-$\ell$ levels ($\ket{52\mrm{D}_{5/2}, m_J = 5/2}$ and $\ket{52\mrm{F}, m_F = 2}$) and circular levels $\ket{n\mrm{C}, 49 \leq n \leq 53}$. The number of detected counts per $\micro\second$ is plotted versus the ionization field (bottom horizontal axis). The upper horizontal axis gives the time delay from the beginning of the ionization ramp. The shaded areas depict the integration windows used for the measurements of the $N_n$s. Each curve is the result of the average over $\approx10^5$ experimental runs.}
\label{fig:ionization} 
\end{figure}

\begin{figure*}[t]
\centering
	\includegraphics[width = \textwidth]{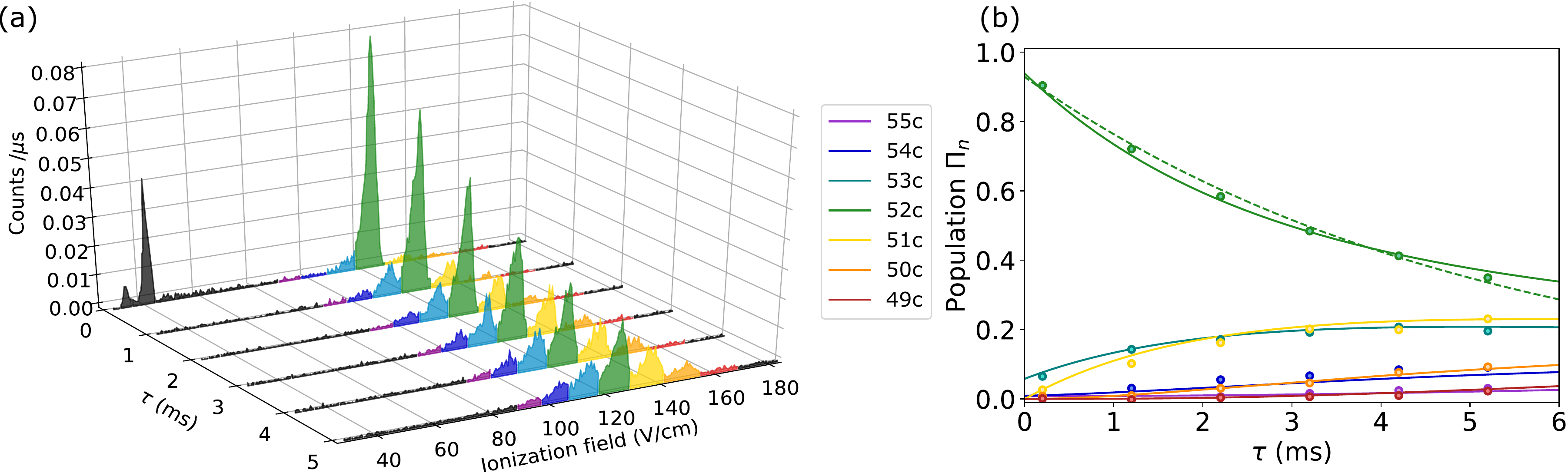}
	\caption{{(a)} Decay of circular Rydberg atoms initially prepared in $\kcirc{52}$. Ionization spectra are plotted as a function of the delay time $\tau$. The color shadings indicate the integration windows for the measurement of the circular level normalized populations, $\Pi_n$s ($49\leq n \leq 55$). Population transfer on the $\kcirc{n}\leftrightarrow\kcirc{(n-1)}$ transitions (frequencies $\nu_{n} = 40.7\,\giga\hertz$, $43.0\,\giga\hertz$, $45.5\,\giga\hertz$, $48.2\,\giga\hertz$, $51.1\,\giga\hertz$, $54.3\,\giga\hertz$ and $57.7\,\giga\hertz$ for $n=55$ to $49$, respectively) is apparent. {(b)} Time evolution of the populations $\Pi_n$s, from the data in (a), as a function of $\tau$. Circles are experimental, solid lines are the results of their best fit to the rate equation model, with $T_{\MW} = 11.1\,\micro\kelvin$. The dashed line is a fit of $\Pi_{52}$ to an exponential decay of time constant $5.1\,\milli\second$. Statistical error bars are smaller than the data points. In (a) and (b), each curve and data point is the result of the average over $\approx10^4$ experimental runs. Panels (a) and (b) and Fig.~\ref{fig:ionization} share the same color code.}
\label{fig:decay} 
\end{figure*}

Figure~\ref{fig:ionization} shows the ionization spectra recorded after the preparation of $\ket{52\mrm{D}_{5/2}, m_J = 5/2}$, $\ket{52\mrm{F}, m=2}$ and $\kcirc{52}$. The ionization ramp is triggered at $t = t_\mrm{ion} = 200\,\micro\second$. It is apparent that the field ionization distinguishes well low- from high-$\ell$ Rydberg levels. We also prepared all circular levels with $48\leq n \leq 53$ by applying $\micro\second$-long single- or multi-photon microwave $\pi$-pulses on the $\kcirc{52}\to\kcirc{n}$ transitions, at frequencies $\nu_\MW= 52.8\,\giga\hertz, 51.2\,\giga\hertz, 49.6\,\giga\hertz, 48.2\,\giga\hertz$ and $45.5\,\giga\hertz$ for $n=48$ to $53$, respectively. The ionization spectra in Fig.~\ref{fig:ionization} reveal that our measurement distinguishes efficiently between circular Rydberg levels with different $n$ values.  

Elliptical levels with high $m$ ($m\lesssim 50$), spuriously prepared by the imperfections of the circularization process, are not addressed by the $\pi$ microwave pulses in the applied electric field. They are left in the $n=52$ manifold when we prepare $\kcirc{(n\neq52)}$ and ionized in nearly the same field as $\kcirc{52}$ ($F = 123\,\volt\per\centi\meter$). From the area of these residual peaks, we estimate a lower bound of the $\kcirc{52}$ purity to be $80\%$. Most of the spurious population is attributed to elliptical levels with $m \geq n-3$, close to the circular state. 
For all relevant circular Rydberg levels $\kcirc{n}$, we determine the total number of detected atoms $N_n$ for each excitation pulse by integrating the ionization signal, corrected for ``dark'' counts ($\approx 0.4 \cdot 10^{-3}\,\mrm{count}\per\micro\second$), over the field windows pictorially shown in Fig.~\ref{fig:ionization}. We observe that, within experimental uncertainties, the detection efficiencies of all these circular Rydberg levels are identical. 

At $T=0\,\kelvin$, the circular state $\kcirc{n}$ decays by spontaneous emission on the single $\kcirc{n} \to \kcirc{(n-1)}$ transition, at frequency $\nu_{n}$, with a free-space rate $\gamma_{n}$ (${\gamma_{52}}^{-1} \approx 35\,\milli\second$). At a finite microwave BBR temperature $T_\MW$, stimulated emission on this transition [rate $n_\mrm{ph}(\nu_{n}) \gamma_{n}$] and absorption on the $\kcirc{n}\to\kcirc{(n+1)}$ transition [rate $n_\mrm{ph}(\nu_{n+1}) \gamma_{n+1}$] reduce the lifetime. Here $n_\mrm{ph}(\nu_{n})$ is the number of BBR photons at frequency $\nu_{n}$. Note that absorption on transitions towards elliptical levels is negligible due to the small corresponding dipole matrix elements~\cite{Gallagher1994}. We have checked by microwave spectroscopy in similar experimental conditions that there is no measurable transfer from circular to elliptical levels over the timescale of our experiments~\cite{Cortinas2019}. Note also that the small fraction of elliptical levels spuriously prepared by the circularization process decays in a similar way as $\kcirc{n}$, with almost the same rates. In the following, we thus analyze the data as if all the atoms were prepared in $\kcirc{52}$, disregarding the small preparation imperfections.

In order to determine the lifetime of the CRAs and $T_\MW$, we prepare the atoms in $\kcirc{52}$ and detect them after a delay time $\tau$ varying from $0.2$ to $5.2\,\milli\second$ (Fig.~\ref{fig:decay}a). The population in the low-$\ell$ levels decays rapidly (lifetime $<200\,\micro\second$~\cite{Beterov2009}) and has completely vanished for $\tau=1.2\,\milli\second$. Population transfers from $\kcirc{n}$ to adjacent circular levels is conspicuous. The total number of atoms in all detected circular levels, $N_\mrm{tot}(\tau)$, is constant, up to experimental drifts of about $6\,\%$, over the whole data set.

In Fig.~\ref{fig:decay}b, we plot the relative populations $\Pi_n(\tau) = N_n(\tau)/N_\mrm{tot}(\tau)$ of the $\ket{n\mrm{C}, 49\leq n \leq 55}$ levels as a function of $\tau$. Strikingly, an exponential fit to the decay of population $\Pi_{52}$ (dashed line) features a time constant of $(5.1\pm 0.2)\,\milli\second$, an exceptionally long time for a Rydberg atom. In a more precise model, the $\Pi_n$s obey a simple rate equation that reads
\begin{align}
	\dot{\Pi}_n &= n_\mrm{ph}(\nu_{n}) \gamma_{n}\ \Pi_{n-1} + [1 + n_\mrm{ph}(\nu_{n+1})] \, \gamma_{n+1}\ \Pi_{n+1} \nonumber \\ 
							& - \{[1 + n_\mrm{ph}(\nu_{n})] \, \gamma_{n} + n_\mrm{ph}(\nu_{n+1}) \, \gamma_{n+1}\} \Pi_n.
\end{align} 
We add to this model a contamination of $\Pi_{n+1}$ by $5\%$ of $\Pi_{n}$ to account for the overlap of the ionization signals. We fit the data points in Fig.~\ref{fig:decay}b with the outcome of the model (solid lines), with $T_\MW$ as the only free parameter. 

The simulation is restricted to values of $n$ between 49 and 55, as $\Pi_{49}$ and $\Pi_{55}$ remain much smaller than one for $\tau\leq 5.2\,\milli\second$. We find a good agreement with our measurements for $T_\MW = (11.1 \pm 0.4) \,\kelvin$, corresponding to $n_\mrm{ph}(\nu_{52}) = (4.3 \pm 0.2)$ (error bars given by the fitting procedure). The lifetime of $\kcirc{52}$ is $(3.7 \pm 0.1)\,\milli\second$, given by the loss rate of $\Pi_{52}$ at $\tau = 0$. Note that the time constant of the simple exponential fit to $\Pi_{52}$ is longer due to the replenishment of $\kcirc{52}$ from the neighboring levels at later times.

The spontaneous-decay rates $\gamma_{n}$ could be affected by the modification of the microwave density of modes due to the surrounding electrodes. By only considering the effect of the closest electrode (the MOT mirror), we find a $20\%$ modification of the rates, which we consider as an upper-bound of the effect of the farther electrodes. Modifying the $\gamma_{n}$s by $\pm20\%$, we find satisfactory fits to the data with $T_\MW=(11\pm2)\,\kelvin$, compatible with the rough estimation given above ($17\,\kelvin$).

\begin{figure}[!tb]
\centering
	\includegraphics[width = 0.85\columnwidth]{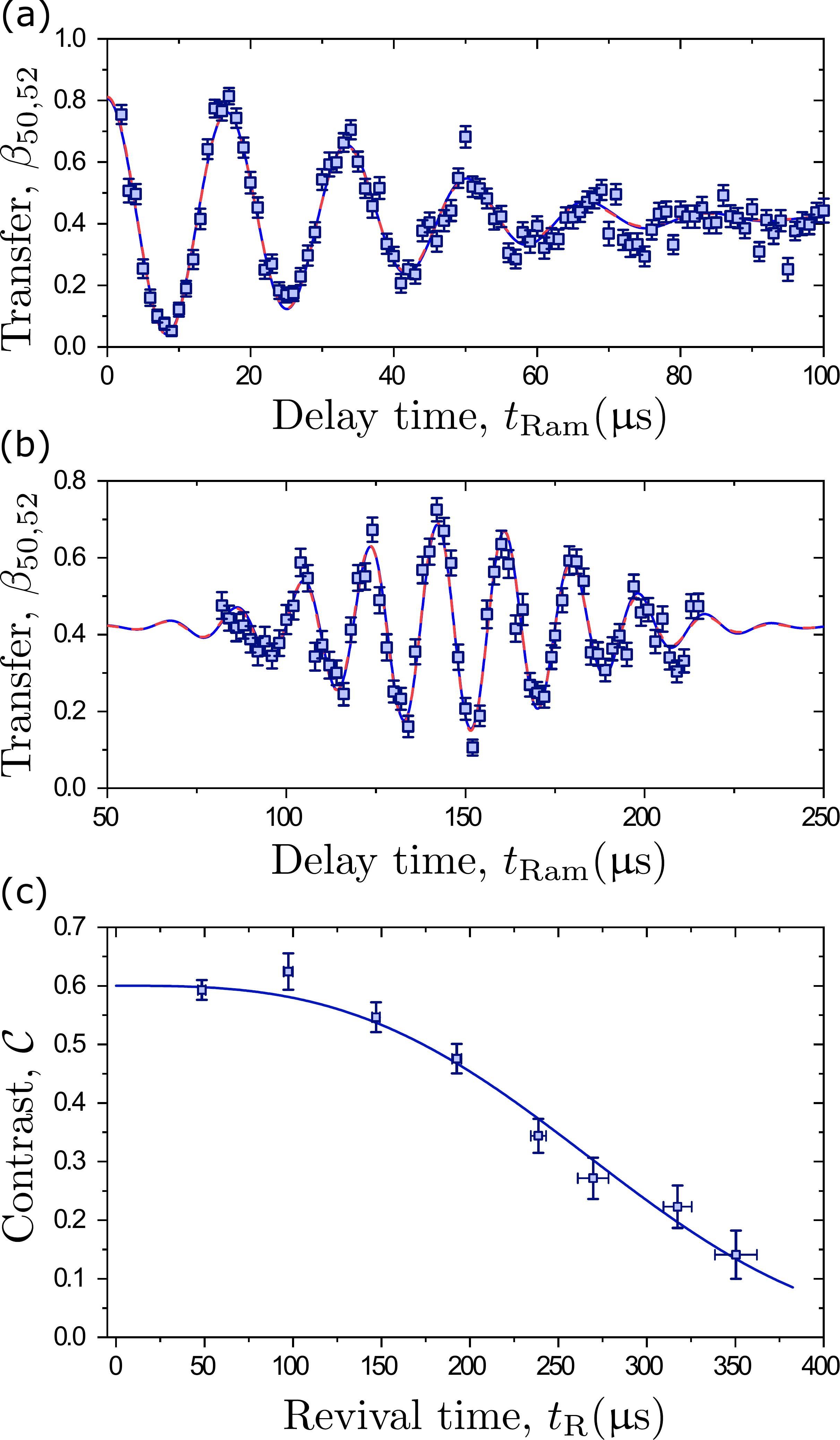}
\caption{{\bf Coherence time measurements.} {(a)} Ramsey fringes and {(b)} Hahn-echo experiment: $\beta_{50,52}$ is plotted w.r.t. the time $\tRam$ between the two $\pi/2$ pulses. In {(b)}, an additional $\pi$ pulse is made at $\tEcho/2 = 75\,\micro\second$ after the first $\pi/2$ pulse. Each point is the result of the average over $5500$ experimental runs. {(c)} Contrast $\mathcal{C}$ of the revival of oscillations of $\beta_{50,52}$ as a function of the revival time $t_\mrm{R}$. In all panels, error bars are statistical. Red dashed lines are fits to a sine with Gaussian damping, blue solid lines are the outcome of the stochastic noise model.}
\label{fig:RamseyEcho}
\end{figure}

We also investigate the coherence properties of the circular Rydberg levels. We record Ramsey interference fringes and Hahn-echo signals on the $\kcirc{52}\to\kcirc{50}$ $(2\times48.2)\,\giga\hertz$-frequency two-photon transition. Here, we reduce the electric field to $F=0.46\,\volt\per\centi\meter$ in order to minimize the sensitivity of the transition to electric field dispersion while keeping a well-defined quantization axis. We then wait $45\,\micro\second$ to let the electric field reach its steady-state, making its residual drifts negligible. 
The microwave source is set $27\,\kilo\hertz$ away from the resonance frequency. We apply two $1.8\,\micro\second$-long $\pi/2$ pulses, separated by a variable waiting time $\tRam$, and measure the fraction $\beta_{50,52} = N_{50}/(N_{50}+N_{52})$ of atoms transferred from $\kcirc{52}$ to $\kcirc{50}$ as a function of $\tRam$ (Fig.~\ref{fig:RamseyEcho}a). For the echo measurements (Fig.~\ref{fig:RamseyEcho}b), we perform an additional $3.2\,\micro\second$-long $\pi$ pulse at a time $\tEcho/2 \leq \tRam$ after the first $\pi/2$ pulse and scan $\tRam$ around $\tEcho$ for 8 different echo times ($50\,\micro\second \leq \tEcho \leq 400\,\micro\second$, $\tEcho = 150\,\micro\second$ in Fig.~\ref{fig:RamseyEcho}b). 
We fit the Ramsey fringes and the Hahn-echo signals to sines sharing identical carrier frequency and width of a Gaussian envelope (red lines in Fig.~\ref{fig:RamseyEcho}). For every dataset, we get a contrast $\mathcal{C}$, \ie, the amplitude of the Gaussian envelope, and a revival time $t_\mrm{R}$, \ie, the position of the maximum of the envelope ($t_\mrm{R}\lesssim \tEcho$). The shared half width at half maximum of the Gaussian envelopes corresponds to a reversible coherence time $\ttwos = (38.5\pm 1)\,\micro\second$.

We plot in Fig.~\ref{fig:RamseyEcho}c the contrast of the echo $\mathcal{C}$ as a function of $t_\mrm{R}$. To reproduce the observed decoherence, we use a simple dephasing model. We consider a Gaussian noise on the energy difference $\Delta E$ between $\kcirc{50}$ and $\kcirc{52}$, of variance ${\sigma_E}^2$, with an exponential correlation function of characteristic time $\tau_\mrm{M}$. Its power spectrum density is Lorentzian with a $3\,\dB$ cut-off frequency of $\nu_\mrm{c} = (2 \pi \tau_\mrm{M})^{-1}$. 
We fit the Ramsey and Hahn-echo signals to this model (blue lines in Fig.~\ref{fig:RamseyEcho}a, b, c), taking as fit parameters $\nu_\mrm{c}$ and $\sigma_E$. It agrees remarkably well with all experimental data, with a cut-off frequency $\nu_\mrm{c} = 76\,\hertz$ and $\sigma_E = h \times 4.7\,\kilo\hertz$. The irreversible decoherence time, $\ttwo$, defined by $\mathcal{C}(\ttwo) = \mathcal{C}(0)/2$, is $\ttwo = 270\,\micro\second$.

The coherence of the CRAs can be limited by both their Stark and Zeeman effects. In order to estimate the contribution of electric field noise or inhomogeneities to the measured $T_2^*$ and $T_2'$ times, we recorded the spectrum of the $\kcirc{52}\to\kell{53}$ transition, where $\kell{53}$ is the low-lying elliptical level with $m = 51$ (Fig.~\ref{fig:setup}b). This $\Delta m = 0$ transition is insensitive to the magnetic field to first order, but its electric field sensitivity, $\alpha_\mrm{CE} = 102\,\mega\hertz\per(\volt\per\centi\meter)$, is large. After an electric-field-gradient minimization, we found a Gaussian line of full width at half maximum (FWHM) $\delta_\mrm{CE} = (175 \pm 9)\,\kilo\hertz$. Correcting for the $121\,\kilo\hertz$ Fourier-limited linewidth of the $10\,\micro\second$-long interrogation pulse, it corresponds to an electric field variation of $\delta F_y = (1.2\pm0.1)\,\milli\volt\per\centi\meter$.

This electric field variation is too small to account for the measured value of $T_2^*$. The $\kcirc{52}\to\kcirc{50}$ two-photon transition has a linear Zeeman effect ($\alpha_\mrm{B} = 2.80\,\mega\hertz\per\Gauss$) but no first order Stark effect. Its quadratic Stark shift is $535\,\kilo\hertz\per(\volt\per\centi\meter)^2$. Small fluctuations $\delta F_y$ around $F_y = 0.46\,\volt\per\centi\meter$ electric field thus result in a frequency shift $\alpha_\mrm{CC} \, \delta F_y$, where $\alpha_\mrm{CC} = 582\,\kilo\hertz\per(\volt\per\centi\meter)$. The measured value of $\delta F_y$ would then induce a reversible coherence time of $4 \ln(2) / (\alpha_\mrm{CC} \delta F_y) = 5\,\milli\second \gg T_2^*$.

The limited decoherence time is thus likely mainly due to magnetic field fluctuations. The fitted value of $\sigma_E$ corresponds to magnetic field fluctuations of $\sigma_E/(h \alpha_\mrm{B}) = 1.7\,\milli\Gauss$. They may result from electric current noise in the MOT magnetic coils, given the $\milli\volt$ noise on the current supply analog control and the circuit bandwidth of about $100\,\hertz$, similar to the estimated $\nu_c$.

We have prepared cold CRAs in a cryogenic environment and measured their lifetime and coherence time. Much longer coherence times, of the order of the circular states lifetime, could be obtained by getting rid of the purely technical magnetic field fluctuations. The $3.7\,\milli\second$ lifetime of $\kcirc{52}$ is already $\approx 40$ times longer than that of low-$\ell$ laser-accessible levels. Combined with ponderomotive laser trapping~\cite{Cortinas2019}, it opens bright perspectives for quantum simulation with circular Rydberg states. The lifetime could even be pushed into the minutes range inside a spontaneous-emission inhibition structure~\cite{Nguyen2018}. Interestingly, recording the transfer of populations between circular Rydberg levels allowed us to estimate \emph{in situ} the \emph{absolute} temperature of the microwave BBR. This could be an important tool for metrology, particularly in the field of atomic clocks, in which the black-body radiation induced shifts significantly contribute to the uncertainty budget~\cite{Levi2010, Ovsiannikov2011, Ushijima2015}.

\medskip
This project has received funding from the European Union's Horizon 2020 research and innovation programme under grant agreement No 817482 (PASQuanS), ERC Advanced grant n\degree\ 786919 (TRENSCRYBE) and QuantERA ERA-NET (ERYQSENS, ANR-18-QUAN-0009-04), from the Region Ile-de-France in the framework of DIM SIRTEQ and from the ANR (TRYAQS, ANR-16-CE30-0026).

\bibliography{bibliography}

\begin{thebibliography}{30}
\expandafter\ifx\csname natexlab\endcsname\relax\def\natexlab#1{#1}\fi
\expandafter\ifx\csname bibnamefont\endcsname\relax
  \def\bibnamefont#1{#1}\fi
\expandafter\ifx\csname bibfnamefont\endcsname\relax
  \def\bibfnamefont#1{#1}\fi
\expandafter\ifx\csname citenamefont\endcsname\relax
  \def\citenamefont#1{#1}\fi
\expandafter\ifx\csname url\endcsname\relax
  \def\url#1{\texttt{#1}}\fi
\expandafter\ifx\csname urlprefix\endcsname\relax\def\urlprefix{URL }\fi
\providecommand{\bibinfo}[2]{#2}
\providecommand{\eprint}[2][]{\url{#2}}

\bibitem[{\citenamefont{Saffman et~al.}(2010)\citenamefont{Saffman, Walker, and
  M\o{}lmer}}]{Saffman2010}
\bibinfo{author}{\bibfnamefont{M.}~\bibnamefont{Saffman}},
  \bibinfo{author}{\bibfnamefont{T.~G.} \bibnamefont{Walker}},
  \bibnamefont{and}
  \bibinfo{author}{\bibfnamefont{K.}~\bibnamefont{M\o{}lmer}},
  \bibinfo{journal}{Rev. Mod. Phys.} \textbf{\bibinfo{volume}{82}},
  \bibinfo{pages}{2313} (\bibinfo{year}{2010}),
  \urlprefix\url{https://link.aps.org/doi/10.1103/RevModPhys.82.2313}.

\bibitem[{\citenamefont{Browaeys and Lahaye}(2020)}]{Browaeys2020}
\bibinfo{author}{\bibfnamefont{A.}~\bibnamefont{Browaeys}} \bibnamefont{and}
  \bibinfo{author}{\bibfnamefont{T.}~\bibnamefont{Lahaye}},
  \bibinfo{journal}{Nat. Phys.} \textbf{\bibinfo{volume}{16}},
  \bibinfo{pages}{132} (\bibinfo{year}{2020}),
  \urlprefix\url{10.1038/s41567-019-0733-z}.

\bibitem[{\citenamefont{Haroche}(2013)}]{Haroche2013}
\bibinfo{author}{\bibfnamefont{S.}~\bibnamefont{Haroche}},
  \bibinfo{journal}{Rev. Mod. Phys.} \textbf{\bibinfo{volume}{85}},
  \bibinfo{pages}{1083} (\bibinfo{year}{2013}),
  \urlprefix\url{https://link.aps.org/doi/10.1103/RevModPhys.85.1083}.

\bibitem[{\citenamefont{Seiler et~al.}(2011)\citenamefont{Seiler, Hogan,
  Schmutz, Agner, and Merkt}}]{Seiler2011}
\bibinfo{author}{\bibfnamefont{C.}~\bibnamefont{Seiler}},
  \bibinfo{author}{\bibfnamefont{S.~D.} \bibnamefont{Hogan}},
  \bibinfo{author}{\bibfnamefont{H.}~\bibnamefont{Schmutz}},
  \bibinfo{author}{\bibfnamefont{J.~A.} \bibnamefont{Agner}}, \bibnamefont{and}
  \bibinfo{author}{\bibfnamefont{F.}~\bibnamefont{Merkt}},
  \bibinfo{journal}{Phys. Rev. Lett.} \textbf{\bibinfo{volume}{106}},
  \bibinfo{pages}{073003} (\bibinfo{year}{2011}),
  \urlprefix\url{https://link.aps.org/doi/10.1103/PhysRevLett.106.073003}.

\bibitem[{\citenamefont{Sedlacek et~al.}(2012)\citenamefont{Sedlacek,
  Schwettmann, K\"ubler, L\"ow, Pfau, and Shaffer}}]{Sedlacek2012a}
\bibinfo{author}{\bibfnamefont{J.~A.} \bibnamefont{Sedlacek}},
  \bibinfo{author}{\bibfnamefont{A.}~\bibnamefont{Schwettmann}},
  \bibinfo{author}{\bibfnamefont{H.}~\bibnamefont{K\"ubler}},
  \bibinfo{author}{\bibfnamefont{R.}~\bibnamefont{L\"ow}},
  \bibinfo{author}{\bibfnamefont{T.}~\bibnamefont{Pfau}}, \bibnamefont{and}
  \bibinfo{author}{\bibfnamefont{J.~P.} \bibnamefont{Shaffer}},
  \bibinfo{journal}{Nat. Phys.} \textbf{\bibinfo{volume}{8}},
  \bibinfo{pages}{819} (\bibinfo{year}{2012}),
  \urlprefix\url{https://www.nature.com/articles/nphys2423}.

\bibitem[{\citenamefont{Facon et~al.}(2016)\citenamefont{Facon, Dietsche,
  Grosso, Haroche, Raimond, Brune, and Gleyzes}}]{Facon2016}
\bibinfo{author}{\bibfnamefont{A.}~\bibnamefont{Facon}},
  \bibinfo{author}{\bibfnamefont{E.-K.} \bibnamefont{Dietsche}},
  \bibinfo{author}{\bibfnamefont{D.}~\bibnamefont{Grosso}},
  \bibinfo{author}{\bibfnamefont{S.}~\bibnamefont{Haroche}},
  \bibinfo{author}{\bibfnamefont{J.-M.} \bibnamefont{Raimond}},
  \bibinfo{author}{\bibfnamefont{M.}~\bibnamefont{Brune}}, \bibnamefont{and}
  \bibinfo{author}{\bibfnamefont{S.}~\bibnamefont{Gleyzes}},
  \bibinfo{journal}{Nature} \textbf{\bibinfo{volume}{535}},
  \bibinfo{pages}{262} (\bibinfo{year}{2016}),
  \urlprefix\url{https://doi.org/10.1038/nature18327}.

\bibitem[{\citenamefont{Cox et~al.}(2018)\citenamefont{Cox, Meyer, Fatemi, and
  Kunz}}]{Cox2018}
\bibinfo{author}{\bibfnamefont{K.~C.} \bibnamefont{Cox}},
  \bibinfo{author}{\bibfnamefont{D.~H.} \bibnamefont{Meyer}},
  \bibinfo{author}{\bibfnamefont{F.~K.} \bibnamefont{Fatemi}},
  \bibnamefont{and} \bibinfo{author}{\bibfnamefont{P.~D.} \bibnamefont{Kunz}},
  \bibinfo{journal}{Phys. Rev. Lett.} \textbf{\bibinfo{volume}{121}},
  \bibinfo{pages}{110502} (\bibinfo{year}{2018}),
  \urlprefix\url{https://link.aps.org/doi/10.1103/PhysRevLett.121.110502}.

\bibitem[{\citenamefont{Graham et~al.}(2019)\citenamefont{Graham, Kwon,
  Grinkemeyer, Marra, Jiang, Lichtman, Sun, Ebert, and Saffman}}]{Graham2019c}
\bibinfo{author}{\bibfnamefont{T.}~\bibnamefont{Graham}},
  \bibinfo{author}{\bibfnamefont{M.}~\bibnamefont{Kwon}},
  \bibinfo{author}{\bibfnamefont{B.}~\bibnamefont{Grinkemeyer}},
  \bibinfo{author}{\bibfnamefont{Z.}~\bibnamefont{Marra}},
  \bibinfo{author}{\bibfnamefont{X.}~\bibnamefont{Jiang}},
  \bibinfo{author}{\bibfnamefont{M.}~\bibnamefont{Lichtman}},
  \bibinfo{author}{\bibfnamefont{Y.}~\bibnamefont{Sun}},
  \bibinfo{author}{\bibfnamefont{M.}~\bibnamefont{Ebert}}, \bibnamefont{and}
  \bibinfo{author}{\bibfnamefont{M.}~\bibnamefont{Saffman}},
  \bibinfo{journal}{Phys. Rev. Lett.} \textbf{\bibinfo{volume}{123}},
  \bibinfo{pages}{230501} (\bibinfo{year}{2019}),
  \urlprefix\url{10.1103/PhysRevLett.123.230501}.

\bibitem[{\citenamefont{Weimer et~al.}(2010)\citenamefont{Weimer, M\"uller,
  Lesanovsky, Zoller, and B\"uchler}}]{Weimer2010}
\bibinfo{author}{\bibfnamefont{H.}~\bibnamefont{Weimer}},
  \bibinfo{author}{\bibfnamefont{M.}~\bibnamefont{M\"uller}},
  \bibinfo{author}{\bibfnamefont{I.}~\bibnamefont{Lesanovsky}},
  \bibinfo{author}{\bibfnamefont{P.}~\bibnamefont{Zoller}}, \bibnamefont{and}
  \bibinfo{author}{\bibfnamefont{H.~P.} \bibnamefont{B\"uchler}},
  \bibinfo{journal}{Nat. Phys.} \textbf{\bibinfo{volume}{6}},
  \bibinfo{pages}{382} (\bibinfo{year}{2010}),
  \urlprefix\url{https://www.nature.com/articles/nphys1614}.

\bibitem[{\citenamefont{Labuhn et~al.}(2016)\citenamefont{Labuhn, Barredo,
  Ravets, de~L\'es\'eleuc, Macr\`i, Lahaye, and Browaeys}}]{Labuhn2016}
\bibinfo{author}{\bibfnamefont{H.}~\bibnamefont{Labuhn}},
  \bibinfo{author}{\bibfnamefont{D.}~\bibnamefont{Barredo}},
  \bibinfo{author}{\bibfnamefont{S.}~\bibnamefont{Ravets}},
  \bibinfo{author}{\bibfnamefont{S.}~\bibnamefont{de~L\'es\'eleuc}},
  \bibinfo{author}{\bibfnamefont{T.}~\bibnamefont{Macr\`i}},
  \bibinfo{author}{\bibfnamefont{T.}~\bibnamefont{Lahaye}}, \bibnamefont{and}
  \bibinfo{author}{\bibfnamefont{A.}~\bibnamefont{Browaeys}},
  \bibinfo{journal}{Nature} \textbf{\bibinfo{volume}{534}},
  \bibinfo{pages}{667} (\bibinfo{year}{2016}),
  \urlprefix\url{https://doi.org/10.1038/nature18274}.

\bibitem[{\citenamefont{Bernien et~al.}(2017)\citenamefont{Bernien, Schwartz,
  Keesling, Levine, Omran, Pichler, Choi, Zibrov, Endres, Greiner
  et~al.}}]{Bernien2017}
\bibinfo{author}{\bibfnamefont{H.}~\bibnamefont{Bernien}},
  \bibinfo{author}{\bibfnamefont{S.}~\bibnamefont{Schwartz}},
  \bibinfo{author}{\bibfnamefont{A.}~\bibnamefont{Keesling}},
  \bibinfo{author}{\bibfnamefont{H.}~\bibnamefont{Levine}},
  \bibinfo{author}{\bibfnamefont{A.}~\bibnamefont{Omran}},
  \bibinfo{author}{\bibfnamefont{H.}~\bibnamefont{Pichler}},
  \bibinfo{author}{\bibfnamefont{S.}~\bibnamefont{Choi}},
  \bibinfo{author}{\bibfnamefont{A.~S.} \bibnamefont{Zibrov}},
  \bibinfo{author}{\bibfnamefont{M.}~\bibnamefont{Endres}},
  \bibinfo{author}{\bibfnamefont{M.}~\bibnamefont{Greiner}},
  \bibnamefont{et~al.}, \bibinfo{journal}{Nature}
  \textbf{\bibinfo{volume}{551}}, \bibinfo{pages}{579} (\bibinfo{year}{2017}),
  \urlprefix\url{https://doi.org/10.1038/nature24622}.

\bibitem[{\citenamefont{Zeiher et~al.}(2017)\citenamefont{Zeiher, Choi,
  Rubio-Abadal, Pohl, van Bijnen, Bloch, and Gross}}]{Zeiher2017}
\bibinfo{author}{\bibfnamefont{J.}~\bibnamefont{Zeiher}},
  \bibinfo{author}{\bibfnamefont{J.-Y.} \bibnamefont{Choi}},
  \bibinfo{author}{\bibfnamefont{A.}~\bibnamefont{Rubio-Abadal}},
  \bibinfo{author}{\bibfnamefont{T.}~\bibnamefont{Pohl}},
  \bibinfo{author}{\bibfnamefont{R.}~\bibnamefont{van Bijnen}},
  \bibinfo{author}{\bibfnamefont{I.}~\bibnamefont{Bloch}}, \bibnamefont{and}
  \bibinfo{author}{\bibfnamefont{C.}~\bibnamefont{Gross}},
  \bibinfo{journal}{Phys. Rev. X} \textbf{\bibinfo{volume}{7}},
  \bibinfo{pages}{041063} (\bibinfo{year}{2017}),
  \urlprefix\url{https://link.aps.org/doi/10.1103/PhysRevX.7.041063}.

\bibitem[{\citenamefont{Hulet et~al.}(1985)\citenamefont{Hulet, Hilfer, and
  Kleppner}}]{Hulet1985}
\bibinfo{author}{\bibfnamefont{R.~G.} \bibnamefont{Hulet}},
  \bibinfo{author}{\bibfnamefont{E.~S.} \bibnamefont{Hilfer}},
  \bibnamefont{and} \bibinfo{author}{\bibfnamefont{D.}~\bibnamefont{Kleppner}},
  \bibinfo{journal}{Phys. Rev. Lett.} \textbf{\bibinfo{volume}{55}},
  \bibinfo{pages}{2137} (\bibinfo{year}{1985}),
  \urlprefix\url{https://link.aps.org/doi/10.1103/PhysRevLett.55.2137}.

\bibitem[{\citenamefont{Saffman and Walker}(2005)}]{Saffman2005}
\bibinfo{author}{\bibfnamefont{M.}~\bibnamefont{Saffman}} \bibnamefont{and}
  \bibinfo{author}{\bibfnamefont{T.~G.} \bibnamefont{Walker}},
  \bibinfo{journal}{Phys. Rev. A} \textbf{\bibinfo{volume}{72}},
  \bibinfo{pages}{022347} (\bibinfo{year}{2005}),
  \urlprefix\url{http://link.aps.org/doi/10.1103/PhysRevA.72.022347}.

\bibitem[{\citenamefont{Corti\~nas et~al.}(2019)\citenamefont{Corti\~nas,
  Favier, Ravon, M\'ehaignerie, Machu, Raimond, Sayrin, and
  Brune}}]{Cortinas2019}
\bibinfo{author}{\bibfnamefont{R.}~\bibnamefont{Corti\~nas}},
  \bibinfo{author}{\bibfnamefont{M.}~\bibnamefont{Favier}},
  \bibinfo{author}{\bibfnamefont{B.}~\bibnamefont{Ravon}},
  \bibinfo{author}{\bibfnamefont{P.}~\bibnamefont{M\'ehaignerie}},
  \bibinfo{author}{\bibfnamefont{Y.}~\bibnamefont{Machu}},
  \bibinfo{author}{\bibfnamefont{J.-M.} \bibnamefont{Raimond}},
  \bibinfo{author}{\bibfnamefont{C.}~\bibnamefont{Sayrin}}, \bibnamefont{and}
  \bibinfo{author}{\bibfnamefont{M.}~\bibnamefont{Brune}},
  \bibinfo{journal}{arXiv:1911.02316}  (\bibinfo{year}{2019}).

\bibitem[{\citenamefont{Assemat et~al.}(2019)\citenamefont{Assemat, Grosso,
  Signoles, Facon, Dotsenko, Haroche, Raimond, Brune, and
  Gleyzes}}]{Assemat2019}
\bibinfo{author}{\bibfnamefont{F.}~\bibnamefont{Assemat}},
  \bibinfo{author}{\bibfnamefont{D.}~\bibnamefont{Grosso}},
  \bibinfo{author}{\bibfnamefont{A.}~\bibnamefont{Signoles}},
  \bibinfo{author}{\bibfnamefont{A.}~\bibnamefont{Facon}},
  \bibinfo{author}{\bibfnamefont{I.}~\bibnamefont{Dotsenko}},
  \bibinfo{author}{\bibfnamefont{S.}~\bibnamefont{Haroche}},
  \bibinfo{author}{\bibfnamefont{J.}~\bibnamefont{Raimond}},
  \bibinfo{author}{\bibfnamefont{M.}~\bibnamefont{Brune}}, \bibnamefont{and}
  \bibinfo{author}{\bibfnamefont{S.}~\bibnamefont{Gleyzes}},
  \bibinfo{journal}{Phys. Rev. Lett.} \textbf{\bibinfo{volume}{123}},
  \bibinfo{pages}{143605} (\bibinfo{year}{2019}).

\bibitem[{\citenamefont{Raimond et~al.}(2010)\citenamefont{Raimond, Sayrin,
  Gleyzes, Dotsenko, Brune, Haroche, Facchi, and Pascazio}}]{Raimond2010}
\bibinfo{author}{\bibfnamefont{J.~M.} \bibnamefont{Raimond}},
  \bibinfo{author}{\bibfnamefont{C.}~\bibnamefont{Sayrin}},
  \bibinfo{author}{\bibfnamefont{S.}~\bibnamefont{Gleyzes}},
  \bibinfo{author}{\bibfnamefont{I.}~\bibnamefont{Dotsenko}},
  \bibinfo{author}{\bibfnamefont{M.}~\bibnamefont{Brune}},
  \bibinfo{author}{\bibfnamefont{S.}~\bibnamefont{Haroche}},
  \bibinfo{author}{\bibfnamefont{P.}~\bibnamefont{Facchi}}, \bibnamefont{and}
  \bibinfo{author}{\bibfnamefont{S.}~\bibnamefont{Pascazio}},
  \bibinfo{journal}{Phys. Rev. Lett.} \textbf{\bibinfo{volume}{105}},
  \bibinfo{pages}{213601} (\bibinfo{year}{2010}),
  \urlprefix\url{http://link.aps.org/doi/10.1103/PhysRevLett.105.213601}.

\bibitem[{\citenamefont{Dietsche et~al.}(2019)\citenamefont{Dietsche, Larrouy,
  Haroche, Raimond, Brune, and Gleyzes}}]{Dietsche2019}
\bibinfo{author}{\bibfnamefont{E.~K.} \bibnamefont{Dietsche}},
  \bibinfo{author}{\bibfnamefont{A.}~\bibnamefont{Larrouy}},
  \bibinfo{author}{\bibfnamefont{S.}~\bibnamefont{Haroche}},
  \bibinfo{author}{\bibfnamefont{J.~M.} \bibnamefont{Raimond}},
  \bibinfo{author}{\bibfnamefont{M.}~\bibnamefont{Brune}}, \bibnamefont{and}
  \bibinfo{author}{\bibfnamefont{S.}~\bibnamefont{Gleyzes}},
  \bibinfo{journal}{Nat. Phys.} \textbf{\bibinfo{volume}{15}},
  \bibinfo{pages}{326} (\bibinfo{year}{2019}),
  \urlprefix\url{https://doi.org/10.1038/s41567-018-0405-4}.

\bibitem[{\citenamefont{Nguyen et~al.}(2018)\citenamefont{Nguyen, Raimond,
  Sayrin, Corti\~nas, Cantat-Moltrecht, Ass\'emat, Dotsenko, Gleyzes, Haroche,
  Roux et~al.}}]{Nguyen2018}
\bibinfo{author}{\bibfnamefont{T.~L.} \bibnamefont{Nguyen}},
  \bibinfo{author}{\bibfnamefont{J.~M.} \bibnamefont{Raimond}},
  \bibinfo{author}{\bibfnamefont{C.}~\bibnamefont{Sayrin}},
  \bibinfo{author}{\bibfnamefont{R.}~\bibnamefont{Corti\~nas}},
  \bibinfo{author}{\bibfnamefont{T.}~\bibnamefont{Cantat-Moltrecht}},
  \bibinfo{author}{\bibfnamefont{F.}~\bibnamefont{Ass\'emat}},
  \bibinfo{author}{\bibfnamefont{I.}~\bibnamefont{Dotsenko}},
  \bibinfo{author}{\bibfnamefont{S.}~\bibnamefont{Gleyzes}},
  \bibinfo{author}{\bibfnamefont{S.}~\bibnamefont{Haroche}},
  \bibinfo{author}{\bibfnamefont{G.}~\bibnamefont{Roux}}, \bibnamefont{et~al.},
  \bibinfo{journal}{Phys. Rev. X} \textbf{\bibinfo{volume}{8}},
  \bibinfo{pages}{011032} (\bibinfo{year}{2018}),
  \urlprefix\url{https://link.aps.org/doi/10.1103/PhysRevX.8.011032}.

\bibitem[{\citenamefont{Goldschmidt et~al.}(2016)\citenamefont{Goldschmidt,
  Boulier, Brown, Koller, Young, Gorshkov, Rolston, and
  Porto}}]{Goldschmidt2016}
\bibinfo{author}{\bibfnamefont{E.~A.} \bibnamefont{Goldschmidt}},
  \bibinfo{author}{\bibfnamefont{T.}~\bibnamefont{Boulier}},
  \bibinfo{author}{\bibfnamefont{R.~C.} \bibnamefont{Brown}},
  \bibinfo{author}{\bibfnamefont{S.~B.} \bibnamefont{Koller}},
  \bibinfo{author}{\bibfnamefont{J.~T.} \bibnamefont{Young}},
  \bibinfo{author}{\bibfnamefont{A.~V.} \bibnamefont{Gorshkov}},
  \bibinfo{author}{\bibfnamefont{S.~L.} \bibnamefont{Rolston}},
  \bibnamefont{and} \bibinfo{author}{\bibfnamefont{J.~V.} \bibnamefont{Porto}},
  \bibinfo{journal}{Phys. Rev. Lett.} \textbf{\bibinfo{volume}{116}},
  \bibinfo{pages}{113001} (\bibinfo{year}{2016}),
  \urlprefix\url{https://link.aps.org/doi/10.1103/PhysRevLett.116.113001}.

\bibitem[{\citenamefont{Zeiher et~al.}(2016)\citenamefont{Zeiher, van Bijnen,
  Schausz, Hild, Choi, Pohl, Bloch, and Gross}}]{Zeiher2016}
\bibinfo{author}{\bibfnamefont{J.}~\bibnamefont{Zeiher}},
  \bibinfo{author}{\bibfnamefont{R.}~\bibnamefont{van Bijnen}},
  \bibinfo{author}{\bibfnamefont{P.}~\bibnamefont{Schausz}},
  \bibinfo{author}{\bibfnamefont{S.}~\bibnamefont{Hild}},
  \bibinfo{author}{\bibfnamefont{J.-y.} \bibnamefont{Choi}},
  \bibinfo{author}{\bibfnamefont{T.}~\bibnamefont{Pohl}},
  \bibinfo{author}{\bibfnamefont{I.}~\bibnamefont{Bloch}}, \bibnamefont{and}
  \bibinfo{author}{\bibfnamefont{C.}~\bibnamefont{Gross}},
  \bibinfo{journal}{Nat. Phys.} \textbf{\bibinfo{volume}{12}},
  \bibinfo{pages}{1095} (\bibinfo{year}{2016}),
  \urlprefix\url{http://dx.doi.org/10.1038/nphys3835}.

\bibitem[{\citenamefont{Teixeira et~al.}(2015)\citenamefont{Teixeira,
  Hermann-Avigliano, Nguyen, Cantat-Moltrecht, Raimond, Haroche, Gleyzes, and
  Brune}}]{Teixeira2015}
\bibinfo{author}{\bibfnamefont{R.~C.} \bibnamefont{Teixeira}},
  \bibinfo{author}{\bibfnamefont{C.}~\bibnamefont{Hermann-Avigliano}},
  \bibinfo{author}{\bibfnamefont{T.~L.} \bibnamefont{Nguyen}},
  \bibinfo{author}{\bibfnamefont{T.}~\bibnamefont{Cantat-Moltrecht}},
  \bibinfo{author}{\bibfnamefont{J.~M.} \bibnamefont{Raimond}},
  \bibinfo{author}{\bibfnamefont{S.}~\bibnamefont{Haroche}},
  \bibinfo{author}{\bibfnamefont{S.}~\bibnamefont{Gleyzes}}, \bibnamefont{and}
  \bibinfo{author}{\bibfnamefont{M.}~\bibnamefont{Brune}},
  \bibinfo{journal}{Phys. Rev. Lett.} \textbf{\bibinfo{volume}{115}},
  \bibinfo{pages}{013001} (\bibinfo{year}{2015}),
  \urlprefix\url{https://link.aps.org/doi/10.1103/PhysRevLett.115.013001}.

\bibitem[{RAM()}]{RAM}
\emph{\bibinfo{title}{{Tessellating Therahertz RAM from Thomas Keating Ltd}}}.

\bibitem[{\citenamefont{Hermann-Avigliano
  et~al.}(2014)\citenamefont{Hermann-Avigliano, Teixeira, Nguyen,
  Cantat-Moltrecht, Nogues, Dotsenko, Gleyzes, Raimond, Haroche, and
  Brune}}]{Hermann-Avigliano2014}
\bibinfo{author}{\bibfnamefont{C.}~\bibnamefont{Hermann-Avigliano}},
  \bibinfo{author}{\bibfnamefont{R.~C.} \bibnamefont{Teixeira}},
  \bibinfo{author}{\bibfnamefont{T.~L.} \bibnamefont{Nguyen}},
  \bibinfo{author}{\bibfnamefont{T.}~\bibnamefont{Cantat-Moltrecht}},
  \bibinfo{author}{\bibfnamefont{G.}~\bibnamefont{Nogues}},
  \bibinfo{author}{\bibfnamefont{I.}~\bibnamefont{Dotsenko}},
  \bibinfo{author}{\bibfnamefont{S.}~\bibnamefont{Gleyzes}},
  \bibinfo{author}{\bibfnamefont{J.~M.} \bibnamefont{Raimond}},
  \bibinfo{author}{\bibfnamefont{S.}~\bibnamefont{Haroche}}, \bibnamefont{and}
  \bibinfo{author}{\bibfnamefont{M.}~\bibnamefont{Brune}},
  \bibinfo{journal}{Phys. Rev. A} \textbf{\bibinfo{volume}{90}},
  \bibinfo{pages}{040502} (\bibinfo{year}{2014}),
  \urlprefix\url{https://link.aps.org/doi/10.1103/PhysRevA.90.040502}.

\bibitem[{\citenamefont{Signoles et~al.}(2017)\citenamefont{Signoles, Dietsche,
  Facon, Grosso, Haroche, Raimond, Brune, and Gleyzes}}]{Signoles2017}
\bibinfo{author}{\bibfnamefont{A.}~\bibnamefont{Signoles}},
  \bibinfo{author}{\bibfnamefont{E.~K.} \bibnamefont{Dietsche}},
  \bibinfo{author}{\bibfnamefont{A.}~\bibnamefont{Facon}},
  \bibinfo{author}{\bibfnamefont{D.}~\bibnamefont{Grosso}},
  \bibinfo{author}{\bibfnamefont{S.}~\bibnamefont{Haroche}},
  \bibinfo{author}{\bibfnamefont{J.~M.} \bibnamefont{Raimond}},
  \bibinfo{author}{\bibfnamefont{M.}~\bibnamefont{Brune}}, \bibnamefont{and}
  \bibinfo{author}{\bibfnamefont{S.}~\bibnamefont{Gleyzes}},
  \bibinfo{journal}{Phys. Rev. Lett.} \textbf{\bibinfo{volume}{118}},
  \bibinfo{pages}{253603} (\bibinfo{year}{2017}),
  \urlprefix\url{https://link.aps.org/doi/10.1103/PhysRevLett.118.253603}.

\bibitem[{\citenamefont{Gallagher}(1994)}]{Gallagher1994}
\bibinfo{author}{\bibfnamefont{T.~F.} \bibnamefont{Gallagher}},
  \emph{\bibinfo{title}{Rydberg {Atoms}}}, Cambridge {Monographs} on {Atomic},
  {Molecular} and {Chemical} {Physics} (\bibinfo{publisher}{Cambridge
  University Press}, \bibinfo{year}{1994}).

\bibitem[{\citenamefont{Beterov et~al.}(2009)\citenamefont{Beterov, Ryabtsev,
  Tretyakov, and Entin}}]{Beterov2009}
\bibinfo{author}{\bibfnamefont{I.~I.} \bibnamefont{Beterov}},
  \bibinfo{author}{\bibfnamefont{I.~I.} \bibnamefont{Ryabtsev}},
  \bibinfo{author}{\bibfnamefont{D.~B.} \bibnamefont{Tretyakov}},
  \bibnamefont{and} \bibinfo{author}{\bibfnamefont{V.~M.} \bibnamefont{Entin}},
  \bibinfo{journal}{Phys. Rev. A} \textbf{\bibinfo{volume}{79}},
  \bibinfo{pages}{052504} (\bibinfo{year}{2009}),
  \urlprefix\url{https://link.aps.org/doi/10.1103/PhysRevA.79.052504}.

\bibitem[{\citenamefont{Levi et~al.}(2010)\citenamefont{Levi, Calosso,
  Calonico, Lorini, Bertacco, Godone, Costanzo, Mongino, Jefferts, Heavner
  et~al.}}]{Levi2010}
\bibinfo{author}{\bibfnamefont{F.}~\bibnamefont{Levi}},
  \bibinfo{author}{\bibfnamefont{C.}~\bibnamefont{Calosso}},
  \bibinfo{author}{\bibfnamefont{D.}~\bibnamefont{Calonico}},
  \bibinfo{author}{\bibfnamefont{L.}~\bibnamefont{Lorini}},
  \bibinfo{author}{\bibfnamefont{E.~K.} \bibnamefont{Bertacco}},
  \bibinfo{author}{\bibfnamefont{A.}~\bibnamefont{Godone}},
  \bibinfo{author}{\bibfnamefont{G.~A.} \bibnamefont{Costanzo}},
  \bibinfo{author}{\bibfnamefont{B.}~\bibnamefont{Mongino}},
  \bibinfo{author}{\bibfnamefont{S.~R.} \bibnamefont{Jefferts}},
  \bibinfo{author}{\bibfnamefont{T.~P.} \bibnamefont{Heavner}},
  \bibnamefont{et~al.}, \bibinfo{journal}{IEEE Trans. Ult. Ferro. Freq. Contr.}
  \textbf{\bibinfo{volume}{57}}, \bibinfo{pages}{600} (\bibinfo{year}{2010}).

\bibitem[{\citenamefont{Ovsiannikov et~al.}(2011)\citenamefont{Ovsiannikov,
  Derevianko, and Gibble}}]{Ovsiannikov2011}
\bibinfo{author}{\bibfnamefont{V.~D.} \bibnamefont{Ovsiannikov}},
  \bibinfo{author}{\bibfnamefont{A.}~\bibnamefont{Derevianko}},
  \bibnamefont{and} \bibinfo{author}{\bibfnamefont{K.}~\bibnamefont{Gibble}},
  \bibinfo{journal}{Phys. Rev. Lett.} \textbf{\bibinfo{volume}{107}},
  \bibinfo{pages}{093003} (\bibinfo{year}{2011}),
  \urlprefix\url{https://link.aps.org/doi/10.1103/PhysRevLett.107.093003}.

\bibitem[{\citenamefont{Ushijima et~al.}(2015)\citenamefont{Ushijima, Takamoto,
  Das, Ohkubo, and Katori}}]{Ushijima2015}
\bibinfo{author}{\bibfnamefont{I.}~\bibnamefont{Ushijima}},
  \bibinfo{author}{\bibfnamefont{M.}~\bibnamefont{Takamoto}},
  \bibinfo{author}{\bibfnamefont{M.}~\bibnamefont{Das}},
  \bibinfo{author}{\bibfnamefont{T.}~\bibnamefont{Ohkubo}}, \bibnamefont{and}
  \bibinfo{author}{\bibfnamefont{H.}~\bibnamefont{Katori}},
  \bibinfo{journal}{Nat. Photon.} \textbf{\bibinfo{volume}{9}},
  \bibinfo{pages}{185} (\bibinfo{year}{2015}).

\end{thebibliography}

\end{document}